\documentclass[conference]{IEEEtran}
\IEEEoverridecommandlockouts
\usepackage{cite}
\usepackage{amsmath,amssymb,amsfonts}
\usepackage{algorithmic}
\usepackage{graphicx}
\usepackage{textcomp}
\usepackage{xcolor}
\def\BibTeX{{\rm B\kern-.05em{\sc i\kern-.025em b}\kern-.08em
    T\kern-.1667em\lower.7ex\hbox{E}\kern-.125emX}}
    \begin{document}
%
\title{ 
Dark Side of HAPS Systems:\\ 
Jamming Threats towards Satellites
}

\author{ Hadil Otay, Khaled Humadi, and Gunes Karabulut Kurt\\
Poly-Grames Research Center, Department of Electrical Engineering\\
Polytechnique Montreal, Montreal, Canada\\
E-mail: hadeeloutay@gmail.com, khaled.humadi@polymtl.ca and gunes.kurt@polymtl.ca
}

\maketitle

\begin{abstract}
Securing satellite communication networks is imperative in the rapidly evolving landscape of advanced telecommunications, particularly in the context of 6G advancements. 
This paper establishes a secure low earth orbit (LEO) satellite network paradigm to address the challenges of the evolving 6G era, with a focus on enhancing communication integrity between satellites and ground stations. Countering the threat of jamming, which can disrupt vital communication channels, is a key goal of this work.
In particular, this paper investigates the performance of two LEO satellite communication scenarios under the presence of jamming attacker. In the first scenario, we consider a system that comprises one transmitting satellite, a receiving ground station, and a high altitude platform station (HAPS) acting as a jammer. The HAPS disrupts communication between the satellite and the ground station, impeding signal transmission. The second scenario involves two satellites, one is the transmitter while the other works as a relay, accompanied by a ground station, and a jamming  HAPS. In this scenario, the transmitting satellite sends data to the ground station using two different paths, i.e.,  direct and indirect transmission paths, with a relay satellite acting as an intermediary in the case of indirect transmission. For both scenarios, we study the system security by developing mathematical frameworks to investigate the outage effect resulting from the jamming signals orchestrated by the HAPS. Our results show that  the satellite cooperation in the second scenario improves the system’s security since the extreme
  jamming effect occurs only when both links are simultaneously disturbed.

\end{abstract}

\section{Introduction}

In recent times, the interest in satellite-based communication systems has surged, driven by industry initiatives from tech giants like SpaceX, Facebook, and Amazon \cite{b1,b2}. This renewed focus extends to satellite-sir-ground-integrated networks (SAGIN), which has prompted research into network design, resource allocation, and performance analysis, highlighting challenges and future directions \cite{b3}. Additionally, researchers and technology providers are shifting their gaze toward the realm of beyond 5G (B5G) and 6G technologies. The potential for 6G technologies, including SAGIN, terahertz communication, orbital angular momentum (OAM), and quantum communication (QC), has drawn attention, leading to proposals for intelligent architectures and enhanced air-to-ground networks \cite{b4}.

While satellite communications (SATCOMs) are pivotal, concerns about their security have grown, particularly regarding the vulnerability of satellite military traffic \cite{b4}. While some anti-spoofing efforts exist, addressing jamming attacks in Low Earth Orbit (LEO) communication for 6G remains a challenge. Traditional anti-jamming techniques, such as modulation and smart antennas, are facing evolving jamming tactics, necessitating innovative solutions. \cite{b5,b6}.


In the realm of secure satellite networks, several articles and surveys have addressed the critical issue of cybersecurity in the emerging space industry. For instance, the authors in \cite{b8} highlight the lack of integrated security measures or outdated security techniques in satellites. Similarly,  \cite{b9} delves into the security issues and vulnerabilities existing in the context of 5G networks.
Although the security issues in 5G overlap with those of 6G, the integration of satellites into 6G networks and the evolving capabilities of hackers pose additional threats to the confidentiality and integrity of satellite communication links. As such several research works, including \cite{b10, b11,b12}, shed light on one aspect of security concerns, 
and provide valuable insights into the impact of jamming attacks on wireless communication systems.

Within the space-air-ground domain, the authors of \cite{b13} define jamming as the transmission of noise at sufficient power within the same frequency band as the transmitter and receiver. This perspective is further supported by the research conducted in \cite{b9} and \cite{b14}. In the realm of satellite networks, the article \cite{b15} identifies two types of jamming attacks: uplink and downlink. Downlink jamming affects SATCOM broadcasts and navigation satellites, while uplink jamming targets payload and command signals. Command signals play a crucial role in satellite missions, as highlighted in \cite{b16}. This article explores the future trends and technical approaches, in particular for satellite  tracking, telemetry, and command systems (TT\&C), which are responsible for transmitting telemetry and telecommand data, as well as determining satellite orbits. The reliability of satellites heavily depend on the performance of the TT\&C system, making it a critical component in the satellite's lifecycle.

In the literature various anti-jamming techniques have been proposed to mitigate the effects of jamming attacks. One notable approach is the use of intelligent reflecting surfaces \cite{b7}. Additionally, game theory-based approaches incorporating deep learning, reinforcement learning, and Stackelberg games have shown promise in combating jamming attacks, as seen in \cite{b17}. Moreover, comprehensive anti-jamming techniques have been explored throughout the entire satellite launch process, as demonstrated in \cite{b6}.



The focal point of this study lies in the exploration of two satellite communication scenarios. The first scenario encompasses a transmitting satellite, a ground station, and a high altitude platform station (HAPS) taking on the role of a jammer. The HAPS, in this context, disrupts the communication that transpires between the satellite and the ground station, thereby causing obstruction in the transmission of signals. In the second scenario, the setup involves a transmitting satellite, a relay satellite, a ground station, and a HAPS, again acting as a source of disruption. Here, the transmitting satellite orchestrates the transmission of signals toward both the relay satellite and the ground station, with the relay satellite acting as an intermediary. Across both scenarios, the deliberate act of jamming executed by the HAPS generates interference that significantly impacts the communication links. For both scenarios, we study the system security by developing mathematical frameworks to investigate the outage effect resulting from the jamming signals orchestrated by the HAPS.


The rest of this paper is organized as follows. In Section II,
 the system model is introduced. Section III is dedicated to the
system performance analysis. In Section IV,
 Numerical results along with necessary discussions are provided. Finally, we conclude this work in Section V.


\section{System model}

\subsection{System architecture}\label{AA}
\begin{figure}[t]
    \centering
\includegraphics[scale=0.26]{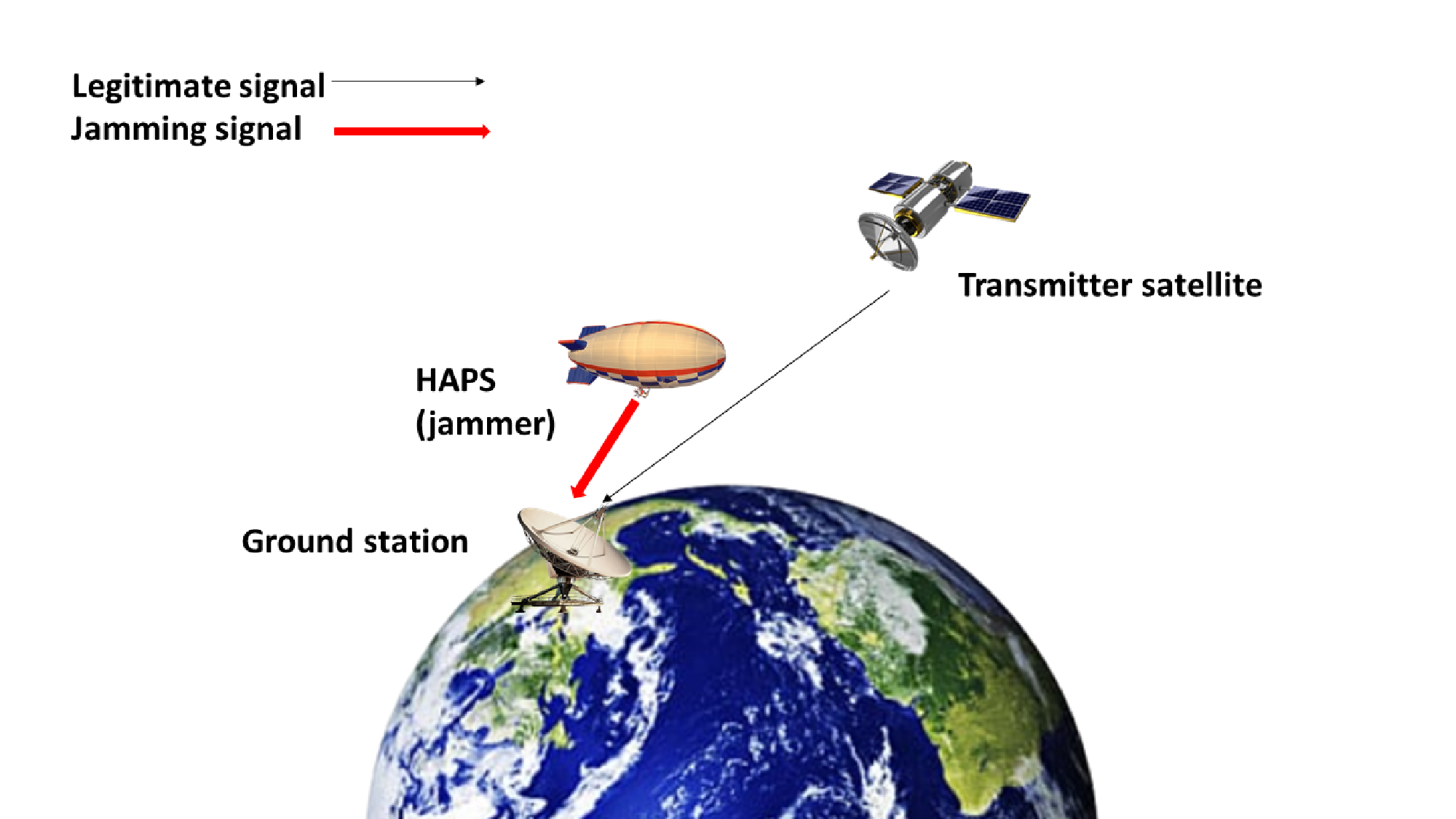}
    \caption{
    System model: Scenario 1. In this scenario, the system comprises a
transmitter satellite (T), a ground station (G), and a HAPS acting
as a jammer.
\vspace{-0.4cm}            }
    \label{sysM1} 
\end{figure}

 We consider a communication system consisting of two distinct scenarios. In the first scenario, the system comprises a transmitter satellite referred to as $T$, a ground station $G$, and a HAPS acting as a jammer. In this configuration, the HAPS interferes with the downlink communication link between the satellite and the ground station, impeding the successful transmission of signals, as illustrated in Fig. \ref{sysM1}.

\begin{figure}[t]
    \centering
    \includegraphics[scale=0.26]{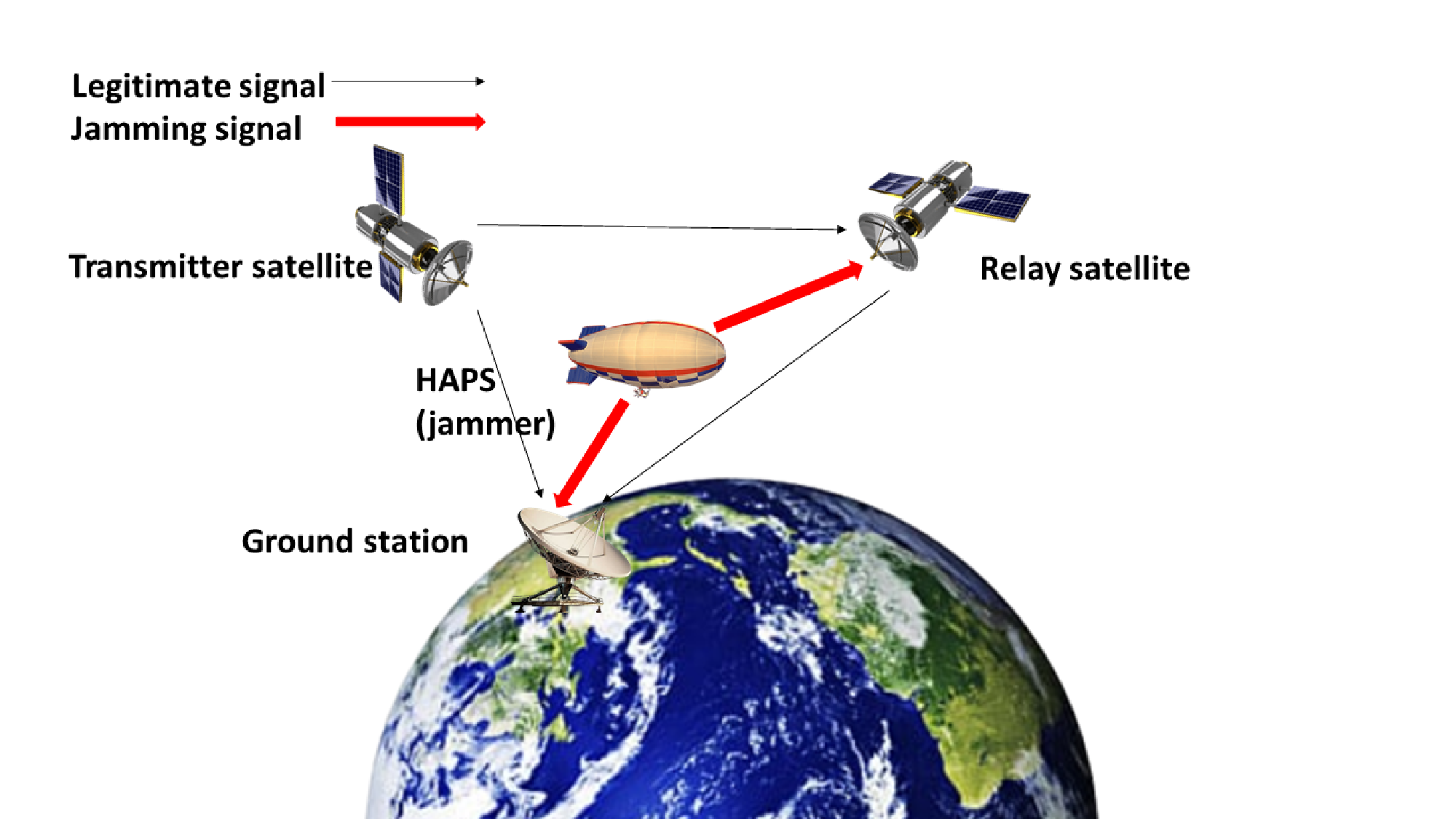}
    \caption{
    System model: Scenario 2. In this scenario, to mitigate the effect of the jamming signal originating from the HAPS, the system comprises cooperation between two 
 satellites (the transmitter (T) and the relay (R)) to send the data of the ground station (G).
 \vspace{-0.4cm}
            }
    \label{sysM2} 
\end{figure}

 In the second scenario, we explore a specific setup involving a transmitter satellite $T$, a relay satellite denoted as $R$, a ground station $G$, and a HAPS that functions as a jammer as well. In this configuration, satellite $T$ transmits signals to satellite $R$, and ground station $G$, simultaneously. The relay satellite $R$ performs its role by forwarding the signals from satellite $T$ to the ground station $G$, thus acting as an intermediary. This enables the ground station to receive its data directly from the satellite $T$ or through the relay satellite $R$ as depicted in Fig. \ref{sysM2}.  Again, the communication links from both satellites $T$ and $R$ to the ground station $G$ are intentionally subjected to disruption through jamming, which is orchestrated by the HAPS.  
   This satellite cooperation is expected to improve the system's security in the addition to the communication performance since the extreme jamming effect  occurs only when both links are simultaneously disturbed.  

\subsection{Channel model}
To analyze the channel characteristics in the different scenarios, we consider both the large-scale fading and the small-scale fading effects. The large-scale fading, which affects the overall link budget, takes into account factors such as path loss, shadowing, and atmospheric attenuation. This enables us to assess the power budget and determine the feasibility of establishing reliable communication links between the satellites and the ground station.

On the other hand, the small-scale fading, specifically Rician fading, is of interest to study the rapid fluctuations in signal amplitude, phase, and power due to multipath propagation, reflections, and interference. By analyzing the Rician fading model, we can understand the statistical behavior of the communication channel, including the presence of a dominant line-of-sight (LOS)  component and the influence of scattered signals. This information aids in designing efficient modulation, coding, and diversity schemes to mitigate the adverse effects of fading and enhance the reliability of the communication links.

\subsubsection{Large-scale fading}
We consider the path loss of the satellites to ground station links as well as the inter-satellite link. The path loss equations can be determined using suitable path loss models that incorporate factors such as distances, frequency, and environmental effects. The inter-satellite path loss between $T$ and $R$ can be expressed as \cite{b19},
\begin{equation}
  PL^{ISL}_{TR} = 32.45 + 20 \textrm{log}_{10} f + 20 \textrm{log}_{10} d_{TR}.   
\end{equation}

On the other hand, the path loss between the transmitter/relay satellite  and \emph{G}  can be expressed in linear scale as \cite{b20}:

\begin{equation}
    PL_{u}=10^{-\big({\frac{PL^{prop}_{u}+PL^{shad}_{u}+PL^{ant}_{u}+PL^{other}_{u}}{10}}\big)},
    \label{largeSF}
\end{equation}
where $u\in\{TG,\; RG\}$ refers to the link from the transmitter \emph{T} or the relay \emph{R} to the ground station \emph{G}.

$PL^{prop}_{u}$ is the path loss in dB, accounting for the free-space propagation between the transmitter/relay satellite and the ground station  calculated as follows
     \begin{equation}
  PL^{prop}_{u}= 32.45 + 20\log_{10} f + 20\log_{10} d_{u}   
\end{equation}

   $PL^{shad}_{u}$  is the shadowing loss in dB, representing the additional signal attenuation due to obstacles and environmental effects between the satellite and the ground station. $PL^{shad}_{u}\sim \mathcal{N}(0,\; \sigma^ 2)$, where $\sigma^ 2$  is the shadowing variance.
   
$PL^{ant}_{u}$  is the antenna gain losses in dB, considering the directional properties of the antennas at the satellite and the ground station, and is calculated as follows \
    \begin{equation}
         PL^{ant}_{u} = 10 \log_{10} \Big(4\Big| \frac{ J_1(2\pi\eta\sin\omega_{u}) }{2\pi\eta\sin\omega_{u}} \Big|^2\Big) 
    \end{equation}
    where $J_1 (.)$  represents the Bessel function of the first kind of order 1, $\eta$ is the aperture radius of the antenna in wavelengths, $\omega_{u}$ is the boresight angle.
    
    $PL^{other}_{u}$ is any other relevant loss, such as polarization loss or losses due to atmospheric effects, between the transmitter or relay satellite and the ground station.


\subsubsection{Small-scale fading}

In our analysis, we make the assumption that the channel conditions remain static over a coherence time interval. This assumption allows us to adopt a flat fading model for the small-scale fading, meaning that the channel response does not vary significantly over the bandwidth of the transmitted signal.

The communication channels between the satellites and the ground station comprise both LOS and  non-LOS (NLOS) components. The likelihood of having a LOS connection increases with the elevation angle, peaking at a 90-degree elevation angle. Taking these factors into account, we model the channel between the satellite and \emph{G} as a Rician fading channel \cite{b18}. This channel fading model is expressed as

\begin{equation}
    h_{u}=\sqrt{{\lambda}} e^{j\phi_{u}}+ \sqrt{\lambda'} h'_{u},
    \label{channelg}
\end{equation}
where $u\in\{TG,\; RG\}$, 
$h'_{u}$ represents the complex channel gain between the transmitter/relay satellite and \emph{G}, and   $\lambda$ is given as 
    
    \begin{equation}
        \lambda=\frac{K_{} }{K_{}+1},
    \end{equation}
 where $K$ is the Rician factor. In (\ref{channelg}), $e^{j\phi_{u}}$ represents the phase of the LOS component, which is a uniform random variable in the range $[-\pi,\pi]$, and
    \begin{equation}
        \lambda' = 1/(K_{}+1) 
    \end{equation}
Here, $\lambda'$ is the power level of the  NLOS link. This Rician fading model accounts for both the deterministic LOS component and the random NLOS component, capturing the effects of multipath propagation and potential scattering between satellites and the ground station.

\section{Performance analysis}

To analyze the system's performance, we adopt the Bernoulli theory, which enables to examine binary events,
namely a non-jammed link and a jammed link. The Bernoulli random variable $X$ is defined based on the comparison of the signal-to-jamming ratio  ($SJR$) to a given threshold $\gamma_{th}$. For the jammed link event, $X$ is set to 1 when $SJR < \gamma_{th}$, representing the occurrence of jamming, whereas, for the non-jammed link event, $X$ is assigned 0 when $SJR\geq \gamma_{th}$, indicating an unhindered communication link.

The $SJR$ captures the power ratio between the received signal and the interfering jamming signal. To compute the $SJR$, we define the received legitimate signal power at the ground station from a given satellite as $Pr_u=Pt_uGt_uGr_uPL_uh_u$, $u\in\{TG,\;RG\}$, where  $Pt_t$, $Gt_u$, and $Gr_u$ denotes the satellite transmit power, satellite transmit antenna gain, and ground station receive antenna gain for path $u$, respectively. The large-scale fading $PL_u$ and small-scale fading $h_u$ are given in (\ref{largeSF}) and (\ref{channelg}), respectively. Similarly, the jamming signal power received from the HAPS  is given as $Pr_j=Pt_{HG}Gt_{HG}Gr_{HG}PL_{HG}h_{HG}$, where $HG$ denotes the link between the HAPS and the ground station G which interfere with both $TG$ and $RG$ links. 
For a given link $u\in\{TG,\;RG\}$, the $SJR$ is expressed as
\begin{equation}
    SJR_u =\frac{Pr_u}{Pr_j}=\frac{Pt_{u} Gt_{u} Gr_{u} PL_u h_u}{Pr_j}.
\end{equation}

To proceed, we need to compute the jamming probability for a given link $u$. This probability is expressed as follows:
\begin{eqnarray}
    \mathbf{Pr}[X = 1]& =& \mathbf{Pr}[SJR_u < \gamma_{th}]\nonumber\\
    &&\!\!\!\!\!\!\!\!\!\!\!= \mathbf{Pr}\bigg[h_u < \frac{\gamma_{th} Pr_j}{Pt_{u} Gt_{u} Gr_{u} PL_u}\bigg],
\end{eqnarray}
while $\mathbf{Pr}[X = 0]=1-\mathbf{Pr}[X = 1]$. 
For the system model in Scenario 1, $SJR=SJR_{TG}$. In the following, we refer to the jamming probability, $P_{jam}(\gamma_{th})=\mathbf{Pr}[SJR < \gamma_{th}]$, where $\gamma_{th}$ is the system reliability threshold. For the analysis tractability, we make some assumptions. First, the Nakagami model is used to approximate the distribution of the Rician random variable.
The PDF distribution of the Nakagami random variable is given as \cite{b23} 
\begin{eqnarray}
f(x)=\frac{\Omega^m}{\Gamma(m)}x^{m-1}e^{-\frac{1}{\Omega}x},\quad\quad x>0
\label{chgain}
\end{eqnarray}
where $m$ and $\Omega$ are the shape and scale parameters, respectively. To match the first and second moments of the Rician and Nakagami distributions, the shape parameter is calculated as $m=(K^2+K+1)/(2K+1)$ which tends to $m = K/2$ for large $K$. Since we consider both LOS and NLOS connections, the channel model in (\ref{chgain}) is rewritten as $f_v(x)=\frac{\Omega_v^{m_v}}{\Gamma(m_v)}x^{m_v-1}e^{-\frac{1}{\Omega_v}x}$, where $v\in\{L,\;N\}$ refers to LOS and NLOS paths. Furthermore, to count for the large-scale fading in the analysis, we consider the propagation path loss since it is the dominant part due to a large distance between satellites and ground stations. As such, we rewrite the path loss as  $PL^{v}_{u} = 32.45 + 20 \log_{10}(f_u) + 10  \alpha^v  \log_{10}(d_u)
$, where  $\alpha^v$ is the path loss exponent and $v\in\{L,\;N\}$ refers to LOS and NLOS paths. The following lemma computes the jamming probability for Scenario 1.

\begin{figure}[t]
    \centering
\includegraphics[scale=0.4]{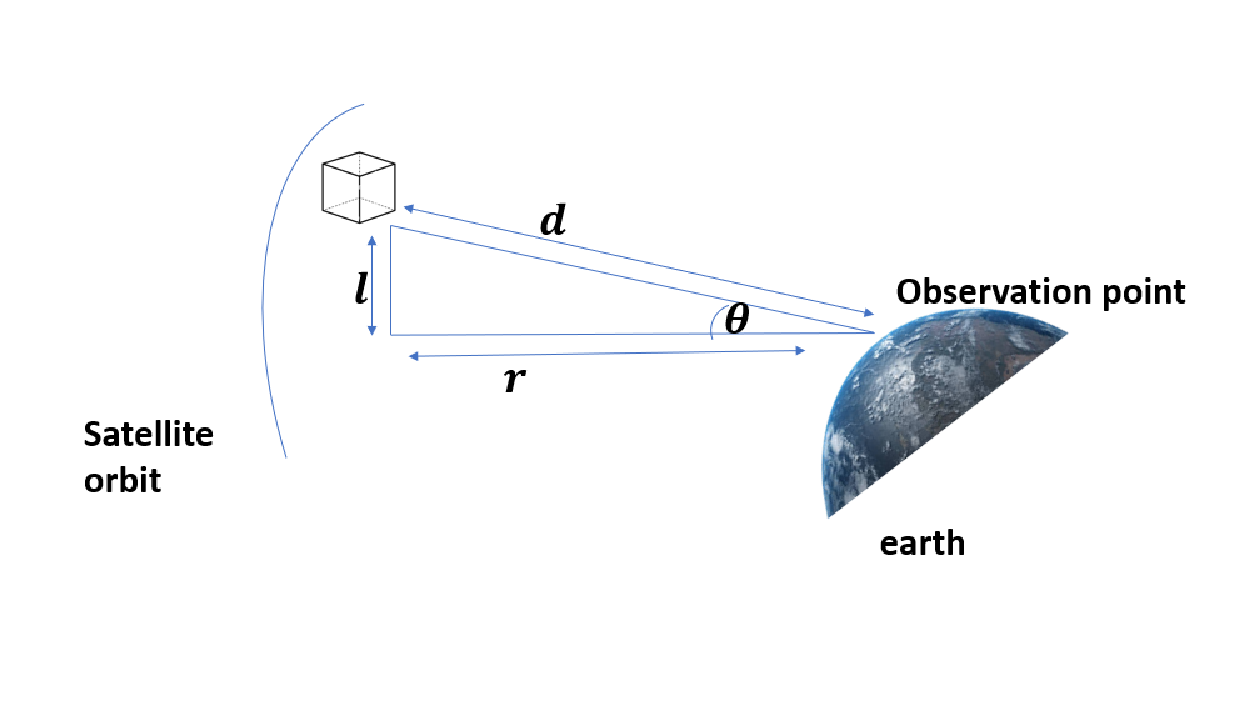}
    \caption{
    The extension of the terrestrial LOS probability model   to satellite
scenarios \cite{b24}.
\vspace{-0.3cm}
}
    \label{LOSFig} 
\end{figure} 

\textbf{Lemma 1:} For a given reliability threshold $\gamma_{th}$, the jamming probability for Scenario 1, denoted as $P^{Sc1}_{jam}(\gamma_{th})$, which is  the  probability that the link \textit{T-G} is jammed by the HAPS, is  expressed as
\begin{eqnarray}
   P^{Sc1}_{jam}(\gamma_{th})&=& P^{TG}_{jam}(\gamma_{th})\nonumber\\
   &&\!\!\!\!\!\!\!\!\!\!\!\!\!\!\!\!\!\!\!\!\!\!\!\!\!\!\!\!\!\!\!\!\!\!\!\!\!\!\!\!\!\!\!\!\!=p_L (\theta_{TG})P^{TG,\;L}_{jam}(\gamma_{th})+p_N (\theta_{TG}) P^{TG,\;N}_{jam}(\gamma_{th})
   \label{covSc1}
\end{eqnarray}
where $p_L(\theta_{TG})$ and $p_N(\theta_{TG})$ define the probability of a link is LOS or NLOS, respectively, between the transmitting satellite and the ground station for a given elevation angle $\theta_{TG}$, $P^{TG,\;L}_{jam}(\gamma_{th})$ and $P^{TG,\;N}_{jam}(\gamma_{th})$ are the conditional jamming probabilities given that the links are in LOS and NLOS, respectively, and are expressed as in (\ref{covLOS}) and (\ref{covNLOS}) at the top of the next page, 
\begin{figure*}
    \begin{eqnarray}
    P^{TG,L}_{jam}(\gamma_{th})&=&\frac{p_L(\theta_{HG})D_L^{TG}}{D^{HG}_L\gamma_{th}}\left(1-\frac{\Omega_L^{m_L}\big(\frac{D_L^{HG} \gamma_{th}}{D_L^{TG}} \Omega_{L}\big)^{-m_L}\Gamma(2m_L) _2F_1\Big(m_L,2m_L,m_L+1,-\frac{D_L^{TG}}{D^{HG}_L\gamma_{th}}\Big)}{\Gamma(m_L)}\right)\nonumber\\
&&\!\!\!\!\!\!\!\!\!\!\!\!\!\!\!\!\!\!\!\!\!\!\!\!\!\!\!\!\!\!\!\!\!\!\!\!\!\!\!\!\!\!\!\!\!\!\!\!+\left(1-\frac{\Omega_N^{m_N}\Big(\frac{D^{HG}_L \gamma_{th}}{D_N^{HG}} \Omega_{L}\Big)^{-m_N}\Gamma(m_L+m_N) _2F_1\Big(m_N,m_L+m_N,m_N+1,-\frac{\Omega_N}{D_L^{TG}\Omega_L D_N^{HG}\gamma_{th}}\Big)}{\Gamma(m_L)}\right)\frac{D_L^{TG}p_N(\theta_{HG})}{D_N^{HG}\gamma_{th}}
\label{covLOS}
\end{eqnarray}
\hrulefill
\vspace{-0.5cm}
\end{figure*}
\begin{figure*}
\begin{eqnarray}
    P^{TG,N}_{jam}(\gamma_{th})&=&\frac{p_L(\theta_{HG})D_N^{TG}}{D^{HG}_L\gamma_{th}}\left(1-\frac{\Omega_L^{m_L}\Big(\frac{D_L^{HG}\gamma_{th}}{D_N^{GT}}  \Omega_{N}\Big)^{-m_L}\Gamma(m_L+m_N) _2F_1\Big(m_L,m_L+m_N,m_L+1,-\frac{\Omega_LD_N^{TG}}{\Omega_N D_L^{HG}\gamma_{th}}\Big)}{\Gamma(m_N)}\right)\nonumber\\
&&\!\!\!\!\!\!\!\!\!+\left(1-\frac{\Omega_N^{m_N}\Big(\frac{D_N^{HG} \gamma_{th}}{D_N^{TG}} \Omega_{N}\Big)^{-m_N}\Gamma(2m_N) _2F_1\Big(m_N,2m_N,m_N+1,-\frac{D_N^{TG}}{D_N^{HG}\gamma_{th}}\Big)}{\Gamma(m_L)}\right)\frac{D_N^{TG}p_N(\theta_{HG})}{D_N^{HG}\gamma_{th}}
\label{covNLOS}
\end{eqnarray}
\hrulefill
\end{figure*}
where, $D_v^{TG}=Pt_{TG} Gt_{TG} Gr_{TG} PL_{TG}^v$ and  $D_v^{HG}=Pt_{HG} Gt_{HG} Gr_{HG} PL_{HG}^v$, $v\in\{L,\;N\}$, refers to the LOS and NLOS link budgets of the useful communication link \textit{T-G} and the jamming link \textit{H-G}, respectively.
\\
\textit{Proof:} see the Appendix

 For the system model in Scenario 2, the ground station receives the same data throughout two different links, i.e., \textit{T-G} and \textit{R-G}.  AS such, the $SJR$ at the ground station is given as
 \begin{eqnarray}
   SJR=\max_{\substack{u\in \{TG,\;RG\}} }\{SJR_u\} 
 \end{eqnarray}
 Accordingly, the jamming probability for Scenario 2 can be obtained using the following lemma.

 \textbf{Lemma 2:} For a given reliability threshold $\gamma_{th}$, the jamming probability for Scenario 2, denoted as $P^{Sc2}_{jam}(\gamma_{th})$, which is  the  probability that both links \textit{T-G} and \textit{R-G} are jammed by the HAPS, is  expressed as
 \begin{eqnarray}
   P^{Sc2}_{jam}(\gamma_{th})&=& P^{TG}_{jam}(\gamma_{th})P^{RG}_{jam}(\gamma_{th}),
   \end{eqnarray}
   where $P^{TG}_{jam}(\gamma_{th})$ is given in (\ref{covSc1}) and $P^{RG}_{jam}(\gamma_{th})$ is given as
   \begin{eqnarray}
   P^{RG}_{jam}(\gamma_{th})&=&\nonumber\\&& \!\!\!\!\!\!\!\!\!\!\!\!\!\!\!\!\!\!\!\!\!\!\!\!\!\!\!\!\!\!p_L (\theta_{RG})P^{RG,\;L}_{jam}(\gamma_{th})+p_N (\theta_{RG}) P^{RG,\;N}_{jam}(\gamma_{th})
\end{eqnarray}

Similarly, $p_L(\theta_{TG})$ and $p_N(\theta_{TG})$ are the probability that  RG link is LOS or NLOS, respectively, for a given elevation angle $\theta_{RG}$, $P^{RG,\;L}_{jam}(\gamma_{th})$ and $P^{RG,\;N}_{jam}(\gamma_{th})$ are the conditional jamming probabilities given that the RG link is in LOS and NLOS, respectively, and are expressed as in (\ref{covLOS}) and (\ref{covNLOS}) with replacing $D_v^{TG}$ with $D_n^{RG}$, $v\in\{L,\;N\}$.

\textit{Proof}: Follows the same steps in the proof of Lemma 1.

To compute $p_L(\theta_{u})$, $u\in\{TG,RG\}$ we apply the model introduced in 
  \cite{b24}, which calculates the link LOS probability between a satellite and a ground station for an arbitrary elevation angle $\theta$ as depicted in Fig. \ref{LOSFig}. 
  This model is expressed as follows
\begin{equation}
p_{L}(\theta) = \exp \left(-\beta \cot(\theta)\right) ,
\end{equation}
where $\beta$ 
is a constant
related to the environment geometry. The elevation angle \( \theta_{} \) can be calculated as $\cot(\theta_{}) = \frac{r}{l}$, 
where \( r \) signifies the horizontal distance between the satellite and the ground station, and \( l \) represents the height of the obstructing structure. 
The probability of the link is NLOS for a given elevation angle $\theta_{}$ is computed at $p_N(\theta_{})=1-p_L(\theta_{})$.

\section{Numerical Results} 

In this section, we provide numerical results to investigate the system security of the two introduced satellite communication scenarios by showing the distribution of the SJR. 
The results offer a comprehensive insight into the interplay of various transmission parameters and angular variations in different environmental scenarios. Unless otherwise specified, the following are the details of the
system parameter. The operating frequency is   2 GHz, transmit powers are $Pt_{TG}=Pt_{RG}=10 $ dB, and $Pt_{HG}=-10$dB. The distances from the ground station to the transmitter and the relay satellites are $d_{TG}=550\times10^{3}$ m and $d_{TG}=600\times10^{3}$ m, respectively, while the distance to the jamming HAPS is $d_{HG}=20\times10^3$ m. The path loss exponents are $\alpha^L=2$, and $\alpha^N=2.2$. The channel fading parameters are set as $m_L=3$, $m_N=2$, $\Omega_L=1/3$, and $\Omega_N=1/2$. To model the LOS probability, three values of $\beta$ are used, i.e., $\beta=\{0.57\;, 0.35\;, 0.048\}$, for suburban, urban, and dense-urban scenarios respectively \cite{b24}. For the jamming signal, we consider the worst case,\textit{ i.e.}, when the HAPS is in LOS with the ground station. In this section, we run Monte Carlo simulations in order to validate the analytical expressions
provided in Lemmas 1 and 2.
\begin{figure}[t]
    \centering
    \includegraphics[scale=0.6]{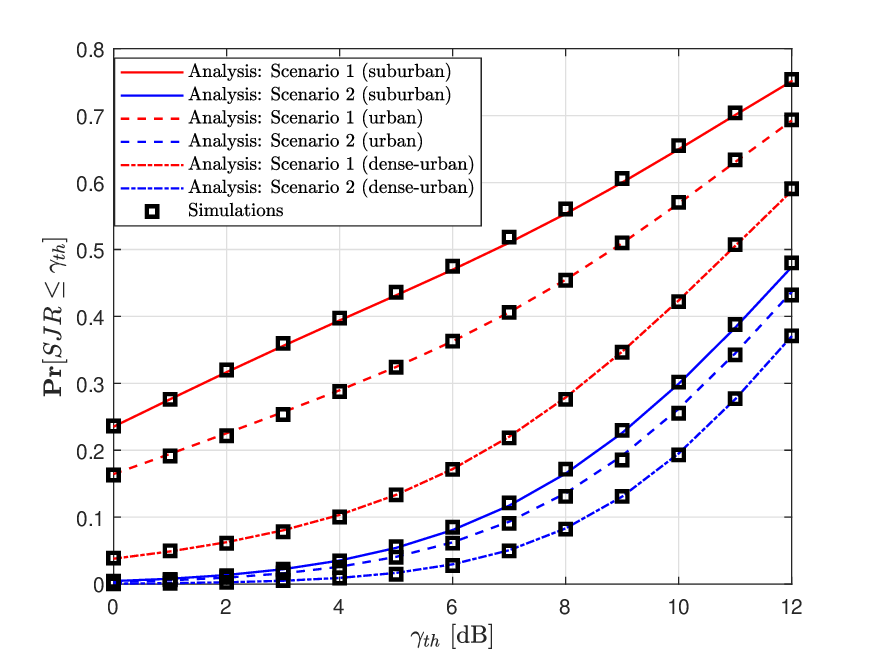}
    \caption{
    CDF of the SJR for Scenarios 1 and 2 with  $\beta=\{0.57\;, 0.35\;, 0.048\}$ for suburban, urban, dense-urban scenarios respectively.
    \vspace{-0.3cm}
    }
    \label{Fig4} 
\end{figure}

In Fig. \ref{Fig4}, we plot the CDF of the $SJR$ as a function of the system reliability threshold for the two scenarios with different values of $\beta$ to consider three different environments, i.e, suburban, urban, and dense-urban areas. As shown in the figure, the analysis results match well with  Simulations which validates our derivations in Section III. Furthermore, this figure shows that for the same system parameters, dense-urban environments provide better SJR performance compared to urban and suburban environments. This is because dense-urban areas result in higher LOS probability for a given elevation angle according to the findings in \cite{b24}. Moreover, Fig. \ref{Fig4} illustrates that the satellite cooperation in Scenario 2 can significantly mitigate the jamming effect compared to the configuration in Scenario 1.     
\begin{figure}[t]
    \centering
    \includegraphics[scale=0.6]{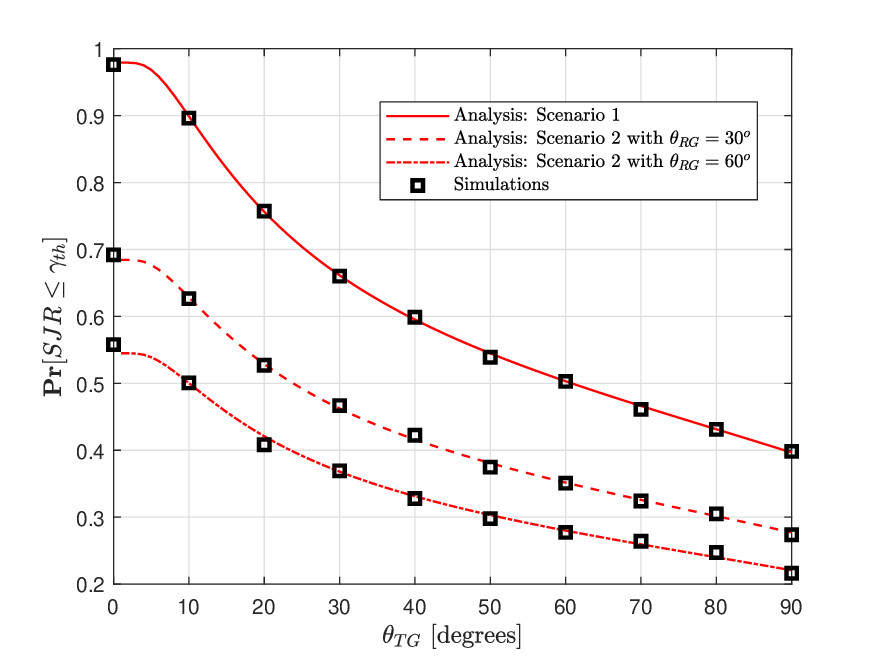}
    \caption{
    CDF of the SJR as a function of the elevation angle of satellite \textit{T} for Scenarios 1 and 2 with $\gamma_{th}=10$ dB, $\beta=0.35$ and with different values of satellite \textit{R} elevation angle $\theta_{RG}$.
    \vspace{-0.3cm}
    }
    \label{Fig5} 
\end{figure}

In Fig. \ref{Fig5}, the CDF of the SJR is plotted as a function of the transmit satellite elevation angle $\theta_{TG}$ with different values of the relay satellite elevation angle, $\theta_{RG}$. As shown in this figure, the system performance of Scenario 2 outperforms that of Scenario 1 at small values of $\theta_{TG}$ while at large values, \textit{i.e}, close to $\theta_{TG}=90^o$ both scenarios provides approximately the same performance. This is encountered due to the fact that at small $\theta_{TG}$, the LOS probability of the TG link is small which degrades the received signal power compared to the jamming signal. As such using a cooperative system as in Scenario 2 can enhance the system's performance.  




\section{Conclusion}                    

 In this paper, we have explored the effectiveness of the indirect link of satellite communication as a robust solution to counter jamming threats. Our findings strongly indicate that the incorporation of a relay satellite represents a viable and effective strategy to combat jamming, ensuring a reliable communication link. In particular, this paper considered a satellite system model for two scenarios: the first scenario uses a single transmit satellite to directly send data to the ground station while the second scenario incorporates a relay system to improve the communication link reliability.  For both
scenarios, we studied system security by developing mathematical frameworks to investigate the outage effect resulting
from the jamming signals orchestrated by an attacker HAPS. Our results showed that using relay satellite systems can provide a safeguarding system against adversarial interference, offering a promising avenue for ensuring dependable and secure communication networks in jamming-prone environments.
\section*{Acknowledgement}
This work is supported in part by Mitacs Globalink.
   \section{Appendix}
\subsection{ Proof of Lemma 1}   
Again the SJR of the TG link is expressed as
\begin{equation}
    SJR_{TG} =\frac{Pr_{TG}}{Pr_j}=\frac{Pt_{TG} Gt_{TG} Gr_{TG} PL_{TG} h_{TG}}{Pr_j}
\end{equation}
where the jamming received power is $Pr_j=Pt_{HG} Gt_{HG} Gr_{HG} PL_{HG} h_{HG}$. In this case, the jamming probability is obtained as 
\begin{eqnarray}
     \mathbf{Pr}[SJR_{TG} < \gamma_{th}]=\mathbf{Pr}\bigg[h_{TG} < \frac{\gamma_{th} Pr_j}{Pt_{TG} Gt_{TG} Gr_{TG} PL_{TG}}\bigg],
\end{eqnarray}
Given the TG is a LOS link, the  jamming probability is written as
\begin{eqnarray}
    P^{TG,L}_{jam}(\gamma_{th})={Pr}\bigg[h^L_{TG} < \frac{\gamma_{th} Pr_j}{D_{L}^{TG}}\bigg],
    \label{eqpr}
\end{eqnarray}
where $D_L^{TG}=Pt_{TG} Gt_{TG} Gr_{TG} PL_{TG}^L$. Since $H_{GT}^L$ follows Nakagami distribution with parameters $m_L$ and $\Omega_L$, then the distribution in (\ref{eqpr}) is expressed as
\begin{eqnarray}
    P^{TG,L}_{jam}(\gamma_{th})&=&\mathbb{E}_{Pr_j}\Bigg[\frac{\Gamma(m_L)-\Gamma\Big(m_L,\;\Omega_L\frac{\gamma_{th} Pr_j}{D_{L}^{TG}}\Big)}{\Gamma(m_L)}\Bigg]\nonumber\\
    &&\!\!\!\!\!\!\!\!\!\!\!\!\!\!\!\!\!\!\!\!\!\!\!\!\!\!\!\!\!\!\!\!\!\!\!\!=1-P_L(\theta_{HG})\mathbb{E}_{h_{HG}^L}\Bigg[\frac{\Gamma\Big(m_L,\;\Omega_L\frac{\gamma_{th} D_{L}^{HG} h_{HG}^L}{D_{L}^{TG}}\Big)}{\Gamma(m_L)}\Bigg]\nonumber\\    &&\!\!\!\!\!\!\!\!\!\!\!\!\!\!\!\!\!\!\!\!\!\!\!\!\!\!\!\!\!\!\!\!\!\!\!\!-P_N(\theta_{HG})\mathbb{E}_{h_{HG}^N}\Bigg[\frac{\Gamma\Big(m_L,\;\Omega_L\frac{\gamma_{th} D_{N}^{HG} h_{HG}^N}{D_{L}^{TG}}\Big)}{\Gamma(m_L)}\Bigg].
    \label{eqjam2}
\end{eqnarray}
Let $\Lambda_L$ and $\Lambda_N$ refer to the second and third terms in (\ref{eqjam2}), respectively. Then,
\begin{eqnarray}   \Lambda_L&=& \frac{\Omega_L^{m_L}}{\Gamma(m_L)}\int_0^\infty \frac{\Gamma\Big(m_L,\;\Omega_L\frac{\gamma_{th} D_{L}^{HG} h_{HG}^L}{D_{L}^{TG}}\Big)}{\Gamma(m_L)} (h_{HG}^L)^{m_L-1}\nonumber\\
&&\times e^{-\frac{1}{\Omega_L}h_{HG}^L}dh_{HG}^L .
\end{eqnarray}
Conducting the above integration and after some manipulations, we get
\begin{eqnarray}
    \Lambda_L&=&\frac{p_L(\theta_{HG})D_L^{TG}}{D^{HG}_L\gamma_{th}}\Bigg(1-\frac{\Omega_L^{m_L}\big(\frac{D_L^{HG} \gamma_{th}}{D_L^{TG}} \Omega_{L}\big)^{-m_L}\Gamma(2m_L) }{\Gamma(m_L)}\nonumber\\
&&\times\frac{_2F_1\Big(m_L,2m_L,m_L+1,-\frac{D_L^{TG}}{D^{HG}_L\gamma_{th}}\Big)}{\Gamma(m_L)}\Bigg) .
\end{eqnarray}
Similarly, $\Lambda_N$ can be derived to be as written in the second term of (\ref{covLOS}). Substituting $\Lambda_L$ and $\Lambda_N$ in (\ref{eqjam2}), we will get the entire expression in (\ref{covLOS}). By following the same steps, we can derive the jamming probability given that the TG link is NLOS to get the results in (\ref{covNLOS}).







\end{document}